\documentstyle[aps, twocolumn]{revtex}

\begin{document}
\title{Comment on paper ``Extremely Low Frequency Plasmons in Metallic
Mesostructures'' [J.B. Pendry, et al., Phys. Rev. Lett. 76, 4773 (1996)]}
\author{Andrey K. Sarychev and Vladimir M. Shalaev}
\address{Department of Physics, New Mexico State University, Las Cruces, NM,
88003}
\maketitle

In a seminal paper \cite{Pendry1996} Pendry, et al. show that in metallic
mesostructures the effective dielectric permittivity $\varepsilon _{e}$ can
be negative, even in the GHz range, which makes possible the excitation of
extremely low-frequency plasmons. The building blocks of the new material
are thin metallic wires with radius $r$ in $\mu {m}$ scale, which are
arranged in a cubic lattice, with lattice constant $a$ in mm scale. The
authors suggest that the effective electron mass can increase by $\sim
10^{4} $ times. This dramatic change in the mass, along with a decrease of
the effective electron density (because of the small fraction of the space
occupied by the wires) result in a depression of the plasma frequency $%
\omega _{p}$ by 6 orders of magnitude. Recently, the negative $\varepsilon
_{e}$ in the GHz range was experimentally verified \cite{Schultz2000}. In
calculations of $\varepsilon _{e}$, Pendry et al. used the Drude formula,
where the original plasma frequency was replaced by its low frequency
counterpart. In the course of calculations, it was implied that the EM
vector potential is constant inside the wires. If the vector potential
inside the wires was ${\bf R}$-dependent, the effective electron mass would
also be ${\bf R}$-dependent and thus, we believe, meaningless.

In our recent papers \cite{SrchvShlv2000PRB} we developed a first-principle
theory for $\varepsilon _{e}$ in mesostructures. Our approach is based on
the direct solution of the Maxwell equations and it does not involve mass
renormalization. For the orthorhombic wire lattice with lattice constants $%
a_{x},$ $a_{y},$ $a_{z}$ and wire radius $r\ll a_{x},$ $a_{y},$ $a_{z}$ we
obtain the dielectric tensor in the following form%
\begin{equation}
\varepsilon _{e,xx}=\varepsilon {_{d}}-\frac{\pi \,r^{2}}{S_{yz}}\frac{\,{{%
\epsilon }_{d}}-[1+\pi \,(S_{yz}/\lambda ^{2})\,{\varepsilon }_{d}]\tilde{%
\varepsilon}{_{m}}}{\,1-\left( {\pi }r/\lambda \right) ^{2}({\cal L}_{yz}+1)%
\tilde{\varepsilon}{_{m}}}  \label{1}
\end{equation}%
where $\lambda $ is the wavelength, $\varepsilon {_{d}}$ is the dielectric
constant of the host material, $S_{yz}=a_{y}a_{z}$, $\tilde{\varepsilon}_{m}{%
=}\varepsilon {_{m}}F(2\pi r\sqrt{\varepsilon {_{m}}}/\lambda )$ is the
renormalized metal permittivity, with $F(x)=2J_{1}\left( x\right) /\left[
xJ_{0}\left( x\right) \right] $ expressed in terms of the Bessel functions $%
J_{0}$ and $J_{1}$, and 
\begin{eqnarray}
{\cal L}_{yz} &=&2\,\log \left( \frac{a_{y}\,\sqrt{1+e^{2}}}{2\,r}\right) + 
\nonumber \\
&&\frac{e\,\pi }{2}-\left( e-\frac{1}{e}\right) \,\arctan e-3,  \label{2}
\end{eqnarray}%
where $e=a_{y}/a_{z}$. Components $\varepsilon _{e,yy}$ and $\varepsilon
_{e,zz}$ of the dielectric tensor can be obtained from Eqs.\thinspace (1)
and (2) by corresponding interchange of indices $x,y,z$. For a cubic
lattice, $a_{x}=a_{y}=a_{z}=a,$ and Eq.\thinspace (1) reduces to 
\begin{equation}
\varepsilon _{e}=\varepsilon {_{d}}-\frac{\pi r^{2}\,\,}{a^{2}}\frac{%
\varepsilon {_{d}}-[1+\pi \left( a/\lambda \right) ^{2}\varepsilon {_{d}]\ }%
\tilde{\varepsilon}{_{m}}}{1-\left( {\pi }r/\lambda \right) ^{2}({\cal L}+1)%
\tilde{\varepsilon}{_{m}}},  \label{3}
\end{equation}%
where ${\cal L}=2\ln (a/\sqrt{2}r)+\pi /2-3$.

In Fig.\thinspace 1 we show the ``loss''-factor $\kappa =|\varepsilon
_{e}^{\prime \prime }/\varepsilon _{e}^{\prime }|$ for $\varepsilon _{e}$
obtained from Eq.\thinspace (3) and from the Drude-like Eq.\thinspace (15)
of \cite{Pendry1996} for $Al$ wire cubic crystal. In Fig.1a, the crystal
parameters are the same as in \cite{Pendry1996}: $r=1{\mu }m$ and $a=5mm$.
The difference between the two approaches, although not dramatic, is still
significant. One might think that for thicker wires, with $r$ much larger
than the penetration (skin) depth $\delta =c/(\omega \,{\rm Im}\sqrt{%
\varepsilon _{m}})$ Pendry's formula works better because the effective
electron mass is nearly ${\bf R}$-independent [for $Al$, $\delta \simeq (2.5/%
\sqrt{\nu }){\mu }m$, with $\nu $ given in ${\rm GHz}$]. However, for $%
r=10\mu {m}$, the difference between the formulae becomes even larger
(Fig.1b). This is because Eq.\thinspace (15) of \cite{Pendry1996} does not
properly take into account the skin effect (which increases with $r$). In
particular, for $\delta /r\ll 1$, Eq.\thinspace (15) predicts that $\kappa
=(\delta /r)^{2}/|\ln (a/r)(1-\omega ^{2}/\omega _{p}^{2})|$, whereas for
strong skin effect the dependence must be $\propto \delta /r$ \cite{LL};
this is the case for Eq.\thinspace (3) giving $\kappa \approx (\delta /r)/|%
{\cal L}(1-\omega ^{2}/\omega _{p}^{2})|$ (for $|\omega -\omega _{p}|\ll
\omega _{p}$). We also note that Eq.\thinspace (3) gives for the ``plasma''
frequency $\omega _{p}^{2}=4c^{2}\pi /(a^{2}{\cal L\,}{{\varepsilon }_{d}})$%
, which differs from the $\omega _{p}$ from Eq.\thinspace (14) of \cite%
{Pendry1996} by a factor of $2\ln (a/r)/({\cal L\,}{{\varepsilon }_{d}})$ [$%
\omega _{p}$ is defined through $\varepsilon _{e}^{\prime }(\omega _{p})=0$
for $\varepsilon _{e}^{\prime \prime }\rightarrow 0$]. The difference in $%
\varepsilon _{e}$ given by Pendry's formula and Eq.\thinspace 3 decreases
with increasing $\delta /r$. However, for $r$ significantly smaller than $%
\delta $ (small skin effect), $\varepsilon _{e}$ is almost purely imaginary,
so that the low-$\omega $ plasmons cannot be excited because of large
losses; also, the vector potential inside the wires, in this case, has a
non-negligible dependence on ${\bf R}$ so that the effective mass of the
electron has questionable meaning. In general, for potential applications of
metal mesostructures it is crucial to have small losses, which is possible
only when the skin effect is large; in this case, we believe, formula (1)\
should be used.{\vspace{-0.1in}}

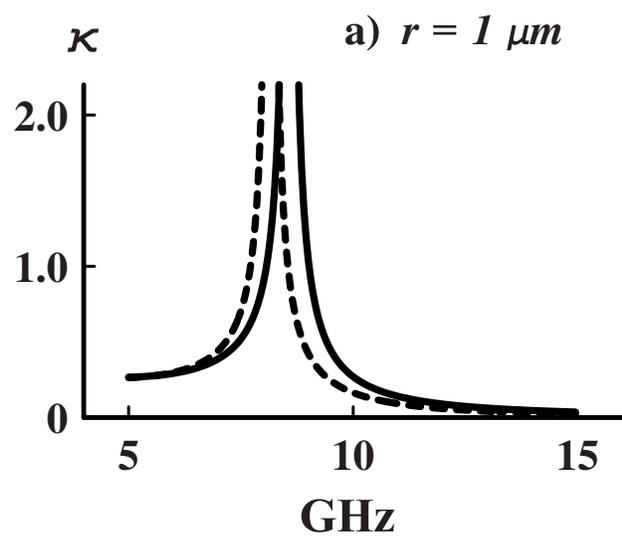
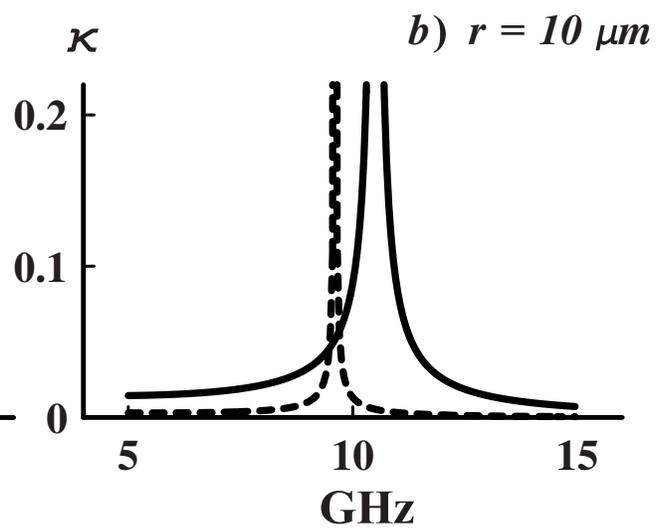
\end{document}